\documentclass[iop]{emulateapj}
\usepackage{apjfonts}

\begin{document}

\title{Two distinct red giant branch populations in the globular cluster NGC~2419 as tracers of a merger event in the Milky Way$^{*}$}
\shorttitle{Two distinct RGBs in NGC~2419}
\shortauthors{Lee et al.}

\author{ Young-Wook Lee$^1$,
Sang-Il Han$^1$,
Seok-Joo Joo$^1$,
Sohee Jang$^1$,
Chongsam Na$^1$,
Sakurako Okamoto$^2$,
Nobuo Arimoto$^{3,4}$,
Dongwook Lim$^1$,
Hak-Sub Kim$^1$,
and Suk-Jin Yoon$^1$}
 
 \altaffiltext{*}{Based on data collected at the Subaru Telescope, which is operated by the National Astronomical Observatory of Japan.}

\affil{$^1$Center for Galaxy Evolution Research and Department of Astronomy, Yonsei University, Seoul 120-749, Korea; ywlee2@yonsei.ac.kr\\
$^2$Kavli Institute for Astronomy and Astrophysics, Peking University, Beijing 100871, China\\
$^3$Subaru Telescope, 650 North Aohoku Place, Hilo, HI 96720, USA\\
$^4$Graduate University for Advanced Studies, 2-21-1 Osawa, Mitaka, Tokyo 181-8588, Japan}

\begin{abstract}
Recent spectroscopic observations of the outer halo globular cluster (GC) NGC 2419 show that it is unique among GCs, in terms of chemical abundance patterns, and some suggest that it was originated in the nucleus of a dwarf galaxy. Here we show, from the Subaru narrow-band photometry employing a calcium filter, that the red giant-branch (RGB) of this GC is split into two distinct subpopulations. Comparison with spectroscopy has confirmed that the redder RGB stars in the $hk$[=(Ca$-b)-(b-y)$] index are enhanced in [Ca/H] by $\sim$0.2 dex compared to the bluer RGB stars. Our population model further indicates that the calcium-rich second generation stars are also enhanced in helium abundance by a large amount ($\Delta$Y = 0.19). Our photometry, together with the results for other massive GCs (e.g., $\omega$~Cen, M22, and NGC~1851), suggests that the discrete distribution of RGB stars in the $hk$ index might be a universal characteristic of this growing group of peculiar GCs. The planned narrow-band calcium photometry for the Local Group dwarf galaxies would help to establish an empirical connection between these GCs and the primordial building blocks in the hierarchical merging paradigm of galaxy formation.
\end{abstract}

\keywords{globular clusters: general --- globular clusters: individual (NGC~2419) --- stars: abundances --- Galaxy: formation}

\section{Introduction}
In the current $\Lambda$CDM hierarchical merging paradigm, a galaxy like the Milky Way formed by numerous mergers of ancient subsystems. Most of the relics of these building blocks, however, are yet to be discovered or identified. The recent discoveries of multiple stellar populations having different heavy element abundances in some massive globular clusters (GCs), such as $\omega$~Centauri, M54, M22, NGC~1851, and Terzan~5 \citep{lee99,bed04,car10a,dac09,jlee09,mar09,han09,car10b,fer09}, are throwing new light on this perspective. The evidence of supernovae (SNe) enrichment in these GCs is indicating that they are most likely the remaining nuclei of more massive primeval dwarf galaxies, rather than being normal GCs. Recent spectroscopy of NGC 2419 \citep{coh10,coh12,muc12} has shown that this outer halo GC was perhaps also able to retain some SNe products (Ca, Sc, K), suggesting that it belongs to this growing group of peculiar GCs. No dispersion in iron is detected, however, from these spectroscopic observations, both from high and low resolution analyses. These observations also found a large spread in K and Mg abundances, bimodality in the abundance distribution of these two elements, and their strong anti-correlation. Nevertheless, the sample size of these spectroscopic observations is rather limited to see whether the metallicity distribution function of RGB stars in this GC is also showing a discrete distribution, which is usually observed in the optical broad-band or calcium narrow-band photometry of other massive GCs suggested to have originated in the nucleus of a dwarf galaxy. The purpose of this Letter is to report our discovery of the two distinct RGB subpopulations in NGC~2419 from the Subaru narrow-band photometry employing a calcium filter.

\begin{figure} 
\includegraphics[scale=0.60]{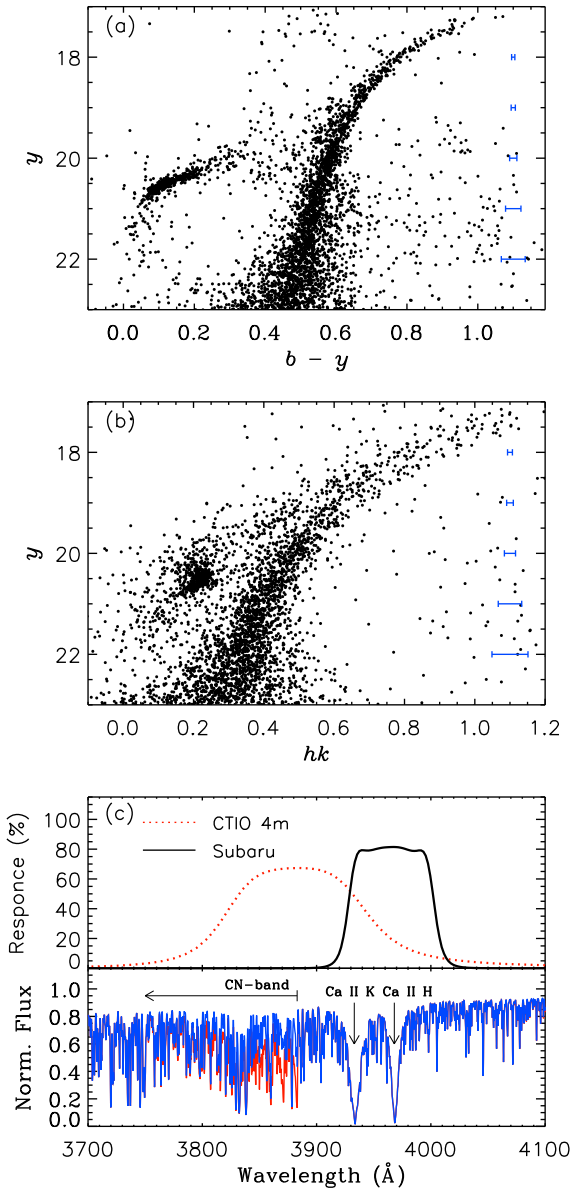}

\caption{CMDs for NGC~2419. The $hk$[=(Ca$-b)-(b-y)$]  index is a measure of calcium abundance \citep{att91}. Note the discrete double RGBs in the ($hk, y$) CMD for the stars brighter than $y$ = 19.5 mag, while this split is not shown in the ($b-y, y$) CMD. The horizontal bars denote the measurement errors ($\pm$1$\sigma$). Panel (c) compares filter transmission functions between the old Ca filter available at CTIO and the new filter employed at Subaru. Synthetic spectra (from \citealt{jlee09}) for CN strong (red line) and CN normal (blue line) stars are also compared to illustrate the regimes occupied by CN band and Ca $_{\rm II}$ H \& K lines. The electronic version of the new Ca filter transmission function is available at our webpage (http://csaweb.yonsei.ac.kr/pub/Ca\_Subaru.txt).}

\label{f:cmd}
\end{figure}

\section{Observations and Color-Magnitude Diagrams}
Our observations in Ca, $b$ and $y$ passbands were performed using the Suprime-Cam \citep{miy02} on the Subaru telescope on the nights of 2012 December 8, 15, and 16. Only the images taken on the photometric condition (mostly from the first night), with the seeing of 0\farcs9 - 1\farcs2, were used. The Suprime-Cam consists of ten 2k $\times$ 4k CCDs providing a pixel scale of 0\farcs2 and covers a 34{\arcmin} $\times$ 27{\arcmin} field of view. The total exposure times for Ca, $b$ and $y$ filters were 14940, 1908, and 954 s, respectively, split into short and long exposures in each band. The pipeline software SDFRED \citep{yag02,ouc04} was used for preprocessing including trimming, bias-subtraction, and flat-fielding. The stand-alone DAOPHOT II, ALLSTAR, and ALLFRAME \citep{ste87,ste94} were used to obtain the point-spread function photometry. The stars in NGC~2420 \citep{att06} were observed during the same nights in order to calibrate instrumental magnitudes to Ca$-by$ standard system. The $hk$ index of \citet{att91}, $hk$ = (Ca$-b)-(b-y)$, was then calculated using our photometry in the Ca, $b$, and $y$ passbands. The old calcium filter available at CTIO, the one used in \citet{jlee09} and \citet{roh11}, was suspected to be deteriorated and affected by CN band at 3883\AA, and this was recently confirmed, upon our request, by C. Johnson, D. H\"olck, and A. Kunder (2012, private communication). The new calcium filter employed in this observation was specially designed by us to avoid this contamination by CN band, and thus it measures practically the H and K lines of singly ionized calcium (see Figure~\ref{f:cmd}c). This new filter set is being used extensively by us in our survey of GCs and dwarf galaxies (see \citealt{joo13}; S. Han et al. 2013, in preparation).

\begin{figure} 
\includegraphics[angle=90,scale=0.4]{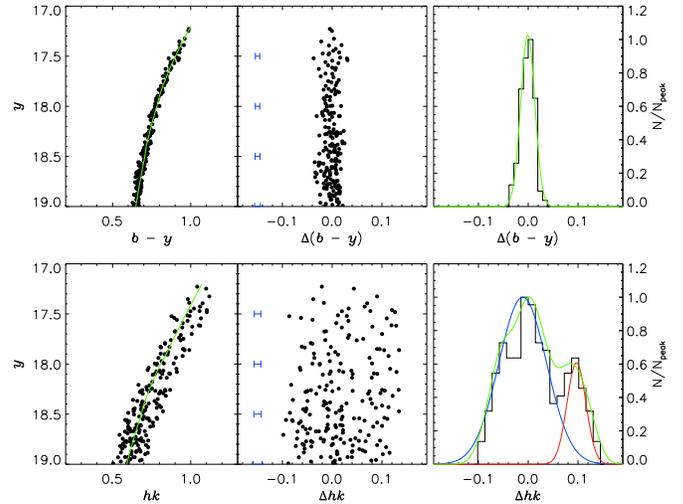}

\caption{Differences in the ($b-y$) color and the $hk$ index, respectively, of each RGB stars from the fiducial lines, the $\Delta$($b-y$) and the $\Delta$$hk$. The blue horizontal bars in the center panels denote measurement errors ($\pm$1$\sigma$) at each magnitude bin. The histograms and the best-fitting Gaussians are plotted in the right panels. Statistical significance for the presence of two subpopulations is more than 99.9\% (see text).}

\label{f:cmdver}
\end{figure}

The color-magnitude diagrams (CMDs) for NGC~2419 in ($b-y, y$) and ($hk, y$) planes are shown in Figure~\ref{f:cmd}. In order to avoid uncertainty caused by possible chip to chip variations, the CMDs in Figure~\ref{f:cmd} were constructed based on the stars only on chip 2, which was placed on the center of the cluster. Furthermore, to examine the CMD features more clearly, ``separation index'' \citep{ste03}, magnitude errors, chi, and sharpness (i.e., $sep$ $\ge$ 1.0, $\sigma$ $\le$ 0.1, $\chi$ $\le$ 1.5, and -0.25 $\le$ SHARP $\le$ 0.25) were used to reject stars affected by blending and adjacent starlight contamination, and those with large photometric uncertainty. The most remarkable feature of Figure~\ref{f:cmd} is the presence of two distinct RGBs in the ($hk, y$) CMD for the RGB stars brighter than y = 19.5 mag, while this split is not shown in the $(b-y, y)$ CMD\footnote[1]{Note that while \citet{dic11} and \citet{bec13} reported color spreads on the RGB of NGC 2419, which are significantly larger than observational errors, the split on the RGB was not detected by their broad-band optical and near $UV$ photometry. However, \citet{bec13} found that blue RGB stars are more centrally concentrated than red RGB stars in $u-V$ color.}. Note that this cannot be due to a differential reddening, because then we would expect even larger separation of the two RGB sequences in the $b-y$ color (see Supplementary Figure 2 of \citealt{jlee09}). In order to examine this feature more clearly, in Figure~\ref{f:cmdver} we have plotted the differences in the ($b-y$) color and the $hk$ index, respectively, of each RGB stars from the fiducial lines, the $\Delta$($b-y$) and the $\Delta$$hk$. The fiducial lines were obtained by connecting the peak (i.e., mode of the distribution) colors at each magnitude bin. The histograms and the best-fitting Gaussians for the RGB stars are plotted in the right panels. For the $\Delta$$hk$, the blue and red Gaussians are obtained from the Gaussian Mixture Modeling (GMM) code \citep{mur10}, while the green line is for the kernel density estimation. The statistical significance for the presence of two subpopulations, based on the GMM likelihood ratio test, is more than 99.9\%. The mean separation between the two RGBs is 0.10 mag in the $hk$ index. The population ratio between the bluer and redder RGB stars is about 0.7:0.3.

\begin{figure} 
\includegraphics[scale=0.45]{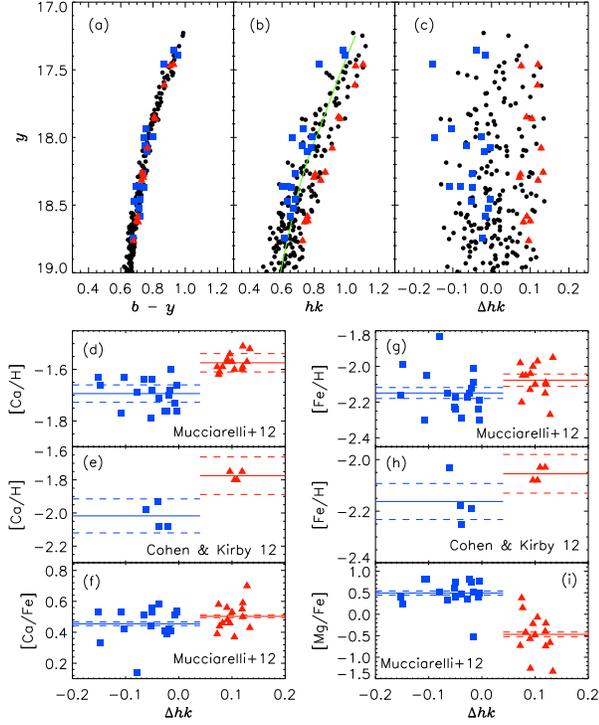}

\caption{Difference in metallicity between the two RGB subpopulations in NGC~2419. Blue squares and red triangles denote blue and red RGB stars in the $\Delta$$hk$ vs. $y$ diagram, respectively, for which cross-matched spectroscopic data are available. It is clear from panels (d) and (e) that the stars having redder $hk$ index are enhanced in Ca abundance than those having bluer index. Panel (f) shows [Ca/Fe] is also different, albeit small, and panels (g) and (h) show a sign of positive correlation between [Fe/H] and $hk$ index. Panel (i) shows Mg deficient stars belong to the subpopulation having redder $hk$ index. The solid and the dashed lines in panels (d) - (i) denote the mean value and $\pm$1$\sigma$ error of the mean value for each group, respectively.}

\label{f:spec}
\end{figure}

\section{DISCUSSION}
We have shown that the RGB of NGC~2419 is split into two distinct subpopulations in the $hk$ versus $y$ CMD. This is most likely indicating the difference in Ca abundance between the two subpopulations, because the Ca filter in the $hk$ index is far more sensitive to changes in Ca abundance than other color indices like $b-y$. In order to confirm this, in Figure~\ref{f:spec}, we have cross-matched our photometry with the spectroscopic data \citep{muc12,coh12} for the relatively bright RGB stars ($y$ ${<}$ 19.0) in the two extreme regimes (i.e., blue and red) on the $\Delta$$hk$ versus $y$ diagram. It is clear from panels (d) and (e) that the RGB stars having redder $hk$ index are indeed enhanced in Ca abundance than those having bluer index. The mean difference between the two subpopulations is 0.12 dex (based on \citealt{muc12}) or 0.24 dex (based on \citealt{coh12}) in [Ca/H], and they are significant at 2.4 and 1.6 $\sigma$ levels, respectively. Furthermore, panel (f) shows [Ca/Fe] is also different, albeit small, by 0.05 dex between the two subpopulations, which is significant at 3.4 $\sigma$ level. Interestingly, panels (g) and (h) show a sign of positive correlation between [Fe/H] and $hk$ index (Ca abundance), although the difference in [Fe/H] is small ($\sim$0.1 dex). These results are significant only at 1.6 and 1.0 $\sigma$ levels, respectively, and further observations are certainly required to confirm this possible difference in [Fe/H]. Panel (i) also shows Mg deficient RGB stars in NGC 2419 belong to the subpopulation having redder $hk$ index. This is consistent with the result by \citet{bec13}, who found that Mg deficient stars are systematically redder in $u-V$ color.

\begin{figure} 
\includegraphics[scale=0.55]{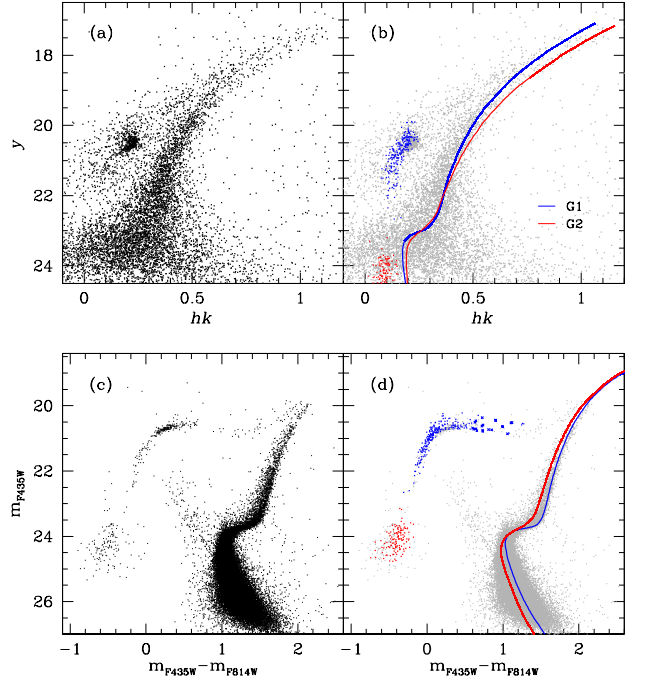}

\caption{Comparison of our population models with the observations for NGC 2419. Observed CMD in panel (c) is by HST ACS/WFC for the F435W and F814W passbands (data from \citealt{dal08}). Models are compared on the observed CMDs in the right panels. Crosses in panel (d) represent RR Lyrae variables. Adopted distance modulus and reddening are $(m-M)$$_{y}$ = 19.88, $E(hk)=-$0.025, and $(m-M$)$_{\rm 435W}$ = 20.03, $E$(F435W-F814W) = 0.105, respectively (see text).}

\label{f:model}
\end{figure}

\begin{table*}
\caption{Input Parameters Adopted in our Best-fit Model of NGC~2419}
\begin{center}
\begin{tabular}{lccccccc}
\hline\hline
Population & Z  & [Fe/H]$^a$  & $\Delta$[CNO/Fe] & Y & Age (Gyr) & Mass Loss (M$_\odot$)$^b$  & Fraction\\
\hline
G1 & 0.00025 & -2.11 & 0.0 & 0.231 & 12.0$\pm$0.3 & 0.179 & 0.7 \\
G2 & 0.00044 & -1.88 & 0.2 & 0.42$\pm$0.04 & 10.0$\pm$0.4 & 0.148 & 0.3\\
\hline
\end{tabular}
\end{center}
$^a${[$\alpha$/Fe]=0.3}\\
$^b${Mean mass loss on the RGB for $\eta$=0.53}\\
\label{t:par}
\end{table*}

Our population models for NGC 2419 are compared with the observed CMDs in Figure 4, and the input parameters adopted in our best-fit models are listed in Table~\ref{t:par}. The latest version of the Yonsei-Yale ($Y^2$) isochrones and horizontal-branch (HB) evolutionary tracks, including the cases of enhanced helium and nitrogen abundances \citep{lee13}, were used in the construction of our population models. Readers are referred to \citet{joo13} for the details of our model construction, and also for the effects of helium and CNO abundances on the CMD morphology. In the $hk$ versus $y$ CMD, the stars in bluer RGB, brighter SGB, and redder HB belong to the metal-poor first-generation subpopulation (G1), while the stars in redder RGB, fainter SGB, and extreme blue horizontal-branch (EBHB) are parts of the metal-rich second-generation subpopulation (G2). In order to reproduce the observed split on the RGB in the $hk$ index, a small, yet important difference in metallicity ($\Delta$[Z/H] = 0.2 dex) is needed, while the presence of EBHB stars in broad-band CMD in lower panels requires a large difference in helium abundance ($\Delta$Y = 0.19 $\pm$ 0.04) between the two subpopulations. \citet{dic11} also suggested a similar difference in helium abundance ($\Delta$Y = 0.16) from their stellar evolution models for NGC 2419. Note that the trend on the RGB is reversed on the broad-band CMD, because, unlike the $hk$ index, the helium effect becomes much more important than the effect caused by a small difference in metallicity in the F435W$-$F814W color. Recent observations have shown that [CNO/Fe] is somewhat enhanced in the metal-rich subpopulation (G2) in most massive GCs \citep{mar12a,mar12b}, and therefore we have further assumed that the G2 is enhanced in [CNO/Fe] by 0.2 dex. In this case, the age difference between the G1 and G2 is predicted to be $\sim$2 Gyr, but this age difference would be reduced to $\sim$1 Gyr, if there were no difference in [CNO/Fe] between the two subpopulations (see \citealt{joo13}). The models presented here are constructed based on the simple assumption that NGC 2419 is composed of only two subpopulations. More detailed modeling, including period-shift analysis of RR Lyrae variables, is required to see whether the G1 population would be further divided into two subpopulations, differing in age, CNO and helium abundances.

The presence of two distinct RGB subpopulations in NGC 2419 is reminiscent of the cases of other massive GCs, such as $\omega$~Cen, M22, and NGC 1851 (\citealt{jlee09}; \citealt{joo13}; S. Han et al. 2013, in preparation), all of which are suggested to have formed in the primeval dwarf galaxy environment. This, in turn, is suggesting that the discrete distribution of RGB stars in the CMD employing the $hk$ index might be a universal characteristic of the remaining nuclei of disrupted dwarf galaxies. The split in the distribution of Ca abundance is most likely due to SNe enrichment. However, unlike other building block candidates, no significant spread in iron is reported in NGC 2419. It also shows an abundance pattern in Mg and K that appears unique among globular clusters and dwarf galaxies \citep{car13,muc12}. Therefore, an alternative scenario, producing the observed K and Mg spread, possibly even a small spread in Ca, and no spread in Fe, has been proposed by \citet{ven12}. Whether the sign of small difference in Fe reported in Figure~\ref{f:spec} is real or not will eventually answer this question, and thus a dedicated spectroscopic observation for more RGB stars in NGC 2419 is urgently required. Note also that the result of our modeling for NGC 2419 is qualitatively similar to those for M22 and NGC 1851 presented by \citet{joo13} in that the G2 is enhanced in both heavy element and helium abundances. This further indicates that NGC 2419 belongs to this group of peculiar GCs, although NGC 2419 appears unique in terms of an abundance pattern in Mg and K. The planned narrow-band photometry employing the same calcium filter for a number of Local Group dwarf galaxies would certainly help to establish an empirical connection between these GCs and the primeval building blocks in the hierarchical merging paradigm of galaxy formation.

\acknowledgments{We thank the referee for a number of helpful suggestions, which led to several improvements in the manuscript. We also thank Emanuele Dalessandro for providing the photometric data of NGC~2419 in machine readable form. Support for this work was provided by the National Research Foundation of Korea to the Center for Galaxy Evolution Research. SJY acknowledges support  from the Mid-career Research Program (No. 2012R1A2A2A01043870) through the NRF of Korea and the DRC program of Korea Research Council of Fundamental Science and Technology (FY 2013).\\}


\begin{thebibliography}{}
\bibitem[Anthony-Twarog et al.(1991)]{att91}Anthony-Twarog, B. J., Laird, J. B., Payne, D., \& Twarog, B. A. 1991, \aj, 101, 1902

\bibitem[Anthony-Twarog et al.(2006)]{att06}Anthony-Twarog, B. J., Tanner, D., Cracraft, M., \& Twarog, B. A. 2006, \aj, 131, 461

\bibitem[Beccari et al.(2013)]{bec13}Beccari, G., Bellazzini, M., Lardo, C., et al. 2013, \mnras, 431, 1995

\bibitem[Bedin et al.(2004)]{bed04}Bedin, L. R., Piotto, G., Anderson, J., et al. 2004, \apj, 605, L125

\bibitem[Carretta et al.(2010a)]{car10a} Carretta, E., Bragaglia, A., Gratton, R. G., et al. 2010a, \apj, 714, L7

\bibitem[Carretta et al.(2010b)]{car10b}Carretta, E., Gratton, R. G., Lucatello, S., et al. 2010b, \apj, 722, L1

\bibitem[Carretta et al.(2013)]{car13}Carretta, E., Gratton, R. G., Bragaglia, A., et al. 2013, \apj, 769, 40

\bibitem[Cohen et al.(2010)]{coh10}Cohen, J. G., Kirby, E. N., Simon, J. D., \& Geha, M. 2010, \apj, 725, 288

\bibitem[Cohen \& Kirby(2012)]{coh12}Cohen, J. G., \& Kirby, E. N. 2012, \apj, 760, 86

\bibitem[Da Costa et al.(2009)]{dac09}Da Costa, G. S., Held, E. V., Saviane, I., \& Gullieuszik, M. 2009, \apj, 705, 1481

\bibitem[Dalessandro et al.(2008)]{dal08}Dalessandro, E., Lanzoni, B., Ferraro, F. R., et al. 2008, \apj, 681, 311

\bibitem[Di Criscienzo et al.(2011)]{dic11}Di Criscienzo, M., Greco, C., Ripepi, V., et al. 2011, \aj, 141, 81

\bibitem[Ferraro et al.(2009)]{fer09}Ferraro, F. R., Dalessandro, E., Mucciarelli, A., et al. 2009, Nature, 462, 483

\bibitem[Han et al.(2009)]{han09}Han, S.-I., Lee, Y.-W., Joo, S.-J., et al. 2009, \apj, 707, L190 


\bibitem[Joo \& Lee(2013)]{joo13}Joo, S.-J. \& Lee, Y.-W. 2013, \apj, 762, 360 

\bibitem[J.-W. Lee et al.(2009)]{jlee09}Lee, J.-W., Kang, Y.-W., Lee, J., \& Lee, Y.-W. 2009, Nature, 462, 480

\bibitem[Lee et al.(1999)]{lee99}Lee, Y.-W., Joo, J.-M., Sohn, Y.-J., et al. 1999, Nature, 402, 55

\bibitem[Lee et al.(2013)]{lee13}Lee, Y.-W., Joo, S.-J., Han, S.-I., et al. 2013, Highlights of Astronomy, 16, in press

\bibitem[Marino et al.(2009)]{mar09}Marino, A. F., Milone, A. P., Piotto, G., et al. 2009, \aap, 505, 1099

\bibitem[Marino et al.(2012a)]{mar12a}Marino, A. F., Milone, A. P., Piotto, G., et al. 2012a, \apj, 746, 14

\bibitem[Marino et al.(2012b)]{mar12b}Marino, A. F., Milone, A. P., Sneden, C., et al. 2012b, \aap, 541, A15

\bibitem[Miyazaki et al.(2002)]{miy02}Miyazaki, S., Komiyama, Y., Sekiguchi, M., et al. 2002, PASJ, 54, 833

\bibitem[Mucciarelli et al.(2012)]{muc12}Mucciarelli, A., Bellazzini, M., Ibata, R., Merle, T., \& Chapman, S. C. 2012, \mnras, 426, 2889

\bibitem[Muratov \& Gnedin(2010)]{mur10}Muratov, A. L. \& Gnedin, O. Y. 2010, \apj, 718, 1266

\bibitem[Ouchi et al.(2004)]{ouc04}Ouchi, M., Shimasaku, K., Okamura, S., et al. 2004, \apj, 611, 660

\bibitem[Roh et al.(2011)]{roh11}Roh, D.-G., Lee, Y.-W., Joo, S.-J., Han, S.-I., Sohn, Y.-J., \& Lee, J.-W. 2011, \apj, 733, L45

\bibitem[Stetson(1987)]{ste87}Stetson, P. B. 1987, PASP, 99, 191

\bibitem[Stetson(1994)]{ste94}Stetson, P. B. 1994, PASP, 106, 250

\bibitem[Stetson et al.(2003)]{ste03}Stetson, P. B., Bruntt, H., \& Grundahl, F. 2003, PASP, 115, 413

\bibitem[Ventura et al.(2012)]{ven12}Ventura, P., DÕAntona, F., Di Criscienzo, M., et al. 2012, \apj, 761, L30

\bibitem[Yagi et al.(2002)]{yag02}Yagi, M., Kashikawa, N., Sekiguchi, M., et al. 2002, \aj, 123, 66

\end{thebibliography}
\end{document}